# Building Free Energy Functional from Atomically-Resolved Imaging: Atomic Scale Phenomena in La-doped BiFeO$_3$

*Anna N. Morozovska[1], Eugene A. Eliseev[2], Deyang Chen[3], Christopher T. Nelson[4], and Sergei V. Kalinin[4*]*

[1] Institute of Physics, National Academy of Sciences of Ukraine,
46, pr. Nauky, 03028 Kyiv, Ukraine

[2] Institute for Problems of Materials Science, National Academy of Sciences of Ukraine,
Krjijanovskogo 3, 03142 Kyiv, Ukraine

[3] Institute for Advanced Materials and Guangdong Provincial Key Laboratory of Optical Information Materials and Technology, South China Academy of Optoelectronics, South China Normal University, Guangzhou 510006, China

[4] The Center for Nanophase Materials Sciences, Oak Ridge National Laboratory, Oak Ridge, TN 37831

**Abstract**

Scanning Transmission Electron Microscopy (STEM) has enabled mapping of atomic structures of solids with sub-pm precision, providing insight to the physics of ferroic phenomena and chemical expansion. However, only a subset of information is available, due to projective nature of imaging in the beam direction. Correspondingly, the analysis often relies on the postulated form of macroscopic Landau-Ginzburg energy for the ferroic long-range order parameter, and some predefined relationship between experimentally determined atomic coordinates and the order parameter field. Here, we propose an approach for exploring the structure of ferroics using reduced order parameter models constructed based on experimental data only. We develop a four sublattices model (FSM) for the analytical description of A-cation displacement in (anti)ferroelectric-antiferrodistortive perovskites of ABO$_3$-type. The model describes the displacements of cation A in four neighboring unit cells and determines the conditions of different structural phases appearance and stability in ABO$_3$. We show that FSM explains the coexistence of rhombohedral, orthorhombic and spatially modulated phases, observed by atomic-resolution STEM in La-doped BiFeO$_3$. Using this approach, we atomically resolve and theoretically model the sublattice asymmetry inherent to the case of the A-site La/Bi cation sublattice in La$_x$Bi$_{1-x}$FeO$_3$ polymorphs. This approach allows exploring the ferroics behaviors from experimental data only, without additional assumptions on the nature of the order parameter.

---

[*] Corresponding author, e-mail: sergei2@ornl.gov

# I. INTRODUCTION

Ferroic materials are the object of continuous fascination for the condensed matter physics community. For over 50 years, the properties of these systems were explored using the combination of scattering techniques that provided the information on the nature and symmetry of corresponding order parameters, and macroscopic property measurements that provided the information on the corresponding expansion coefficients and the nature of phase transitions [1-7]. Once available, the free energy expansion in powers of order parameter(s) is employed in Landau-Ginsburg-Devonshire (**LGD**) free energy [8], and can further be used in the phase field modeling of macro- and nanosized ferroelectrics [9]. Obviously, the nature of boundary conditions at surfaces and interfaces were typically postulated, in the form of (poorly known) correlation and screening lengths [8-9, 14]. Consequently, this approach worked relatively poorly for systems such as polar nanoregions and nanodomains in relaxor ferroelectrics [1], morphotropic systems, or the atomic-scale alternation of polarization in antiferroelectrics and modulated phases [2].

Understanding of ferroic behavior at surfaces, interfaces, and defects as well as the nature of ferroelectric states, considerably advanced in last decades, with the advancement of (Scanning) Transmission Electron Microscopy (shortly **(S)TEM**) [3 – 6]. Probing the unit-cell level symmetry breaking via STEM allowed the determination of direct atomic positions [4,10-12], from which the spatial distributions of order parameter fields can be mapped. However, these analyses to date have been based on two fundamental assumptions. Namely, the nature of the order parameter was assumed to be that of one of the bulk phases of the material. Secondly, the relationship between the experimentally measured atomic coordinates and the order parameter was postulated via certain ad-hoc model [13-14].

Here we derive a model LGD-type free energy describing directly observable degrees of freedom available from atomic-resolution STEM. We propose the theoretical model of four sublattices (shortly **FSM**) for the analytical description of cation displacement in (anti)ferroelectric-antiferrodistortive perovskites of $ABO_3$-type, that explain the coexistence of rhombohedral (**R**), orthorhombic (**O**) and spatially modulated (**SM**) phases observed by atomic-resolution STEM. Using this approach, we atomically resolve and theoretically model the sublattice asymmetry inherent to the case of the A-site La/Bi cation sublattice in perovskite $La_xBi_{1-x}FeO_3$ polymorphs.

## II. EXPERIMENTAL RESULTS

As a model system, we use bismuth ferrite (**BFO**) solid solution. BFO itself is a multiferroic with high ferroelectric Curie temperature $T_C$=1100K and antiferromagnetic Neel temperature $T_N$=650K, high remanent ferroelectric polarization (~90μC/cm$^{-2}$) along [111] axis and antiferromagnetic order coexisting at room temperatures [15,16]. In addition to the rhombohedral (R) *R3c* host phase [17], there are numerous polymorphs experimentally identified in BFO, including epitaxial strain stabilized ferroelectric tetragonal [18], monoclinic [19], and orthorhombic (O) phases [20], as well as a rare-earth dopant stabilized orthorhombic *Pbam* or *Pnma* phases of antiferroelectric type [21-26] (such as in $PbZrO_3$), which forms a high piezoelectric response at morphotropic phase boundary (**MPB**) [23].

Here we explore the MPB of $La_xBi_{1-x}FeO_3$ (**BFO:La**), when 17%La doping stabilizes a phase coexistence between the host ferroelectric rhombohedral and orthorhombic phases at room temperature. The precise space group is not elucidated here but our prior experiments showed projected atomic positions in STEM images isostructural to *Pbam* $PbZrO_3$ as well as antiferroelectric behavior by P-E double loops [21]. Both of these phases exhibits a large principle displacive polar distortion of the La/Bi A-site from the pseudocubic position; for the O phase this consists in the in-plane displacements on alternating pairs of [101]$_{pseudocubic}$ planes in a ++ − − ++… type pattern (**Fig. 1a**), whereas for the ferroelectric *R3c* phase this consists of displacements along the [111] polarization direction resulting in a uniformly polarized ++++… pattern (**Fig. 1b**).

Here, $La_{0.17}Bi_{0.83}FeO_3$ thin films were fabricated on $SrRuO_3$/$SrTiO_3$ sub-layer deposited on buffered Si substrates using pulsed laser deposition. The films exhibit coexistence of ferroelectric *R3c* (**Fig 1a**) and antiferroelectric O-phases (**Fig 1b**) with phase boundaries forming preferentially on [101] planes [21]. STEM images centered at one such boundary are shown for the high-angle annular dark-field (HAADF), and annular bright field (ABF) detectors (see **Fig 1c** and **Fig 1d**, respectively), the former with bright atom contrast sensitive to the atom column Z, and the latter an approximately dark atom contrast image with higher sensitivity to light elements like oxygen. The difference in structure, especially the A-site sublattice displacements (**Fig 1e**) is readily apparent, even from the raw HAADF and ABF images. The [101] plane bisecting the figure contains both the alternating +[101] and –[101] directions of the A-site distortions of the *Pbam* phase, as well as the [−1,±1,−1] direction of the *R3c* phase distortions. In this manner the boundary mimics the local antiferroelectric distortion inherent to the *Pbam* phase. From the [101] displacement statistics vs. distance normal to the interface (**Fig f**) the interface appears atomically sharp within experimental error bars.

The dataset in **Fig 1e** was derived from a single HAADF image and, therefore, subject to scanning artifacts from positional drift during the slow-raster of the electron probe over the area of interest. The result is significantly higher error in relative positions for vertically offset features (the slow-scan axis) compared horizontal features. As a result, the image was corrected along the vertical axis for dilation and shear measured as x-axis correlated variation of atomic spacing from the global mean. XY positions of A site and B site cations were determined by Gaussian fit. Y-axis corrections were smoothed (via a spline fit) and the local transform was applied as a best fit to minimize the variation from the global mean.

Foremost, the high atomic numbers Z of the A-site cations ($Z_{Bi}$ = 83, $Z_{La}$ = 57) compared to B-site cations ($Z_{Fe}$=26) and oxygen ($Z_O$=8), lead to the dominant contribution of A site to the HAADF signal, and thus they exhibit a much higher signal to noise ratio and lower atom positioning error [12]. Moreover, assuming chemical homogeneity on the A-site columns, the probe incident on adjacent A-sites exhibits similar scattering environments with respect to channeling, etc. The result is that experimental measurements of the A-site in isolation have considerably smaller error and greater robustness against artifacts such as from off axis tilt [27] compared to positional non-centrosymmetry analysis that also incorporates Fe or O sublattices. There is a potential point of uncertainty in utilizing the isolated A-site sublattice as in some cases the centrosymmetric reference point can be ill-defined. If, for instance, the images in **Fig.1** contain only a uniform *R3c* phase, from the A-site positions alone the displacement magnitudes are unknown, and so the ++++ ferroelectric and 0000 paraelectric phases cannot be distinguished. Thus, the B-cation sublattice is useful for establishing a reference lattice to measure the A-site and was the method used for the dataset in **Fig 1e**.

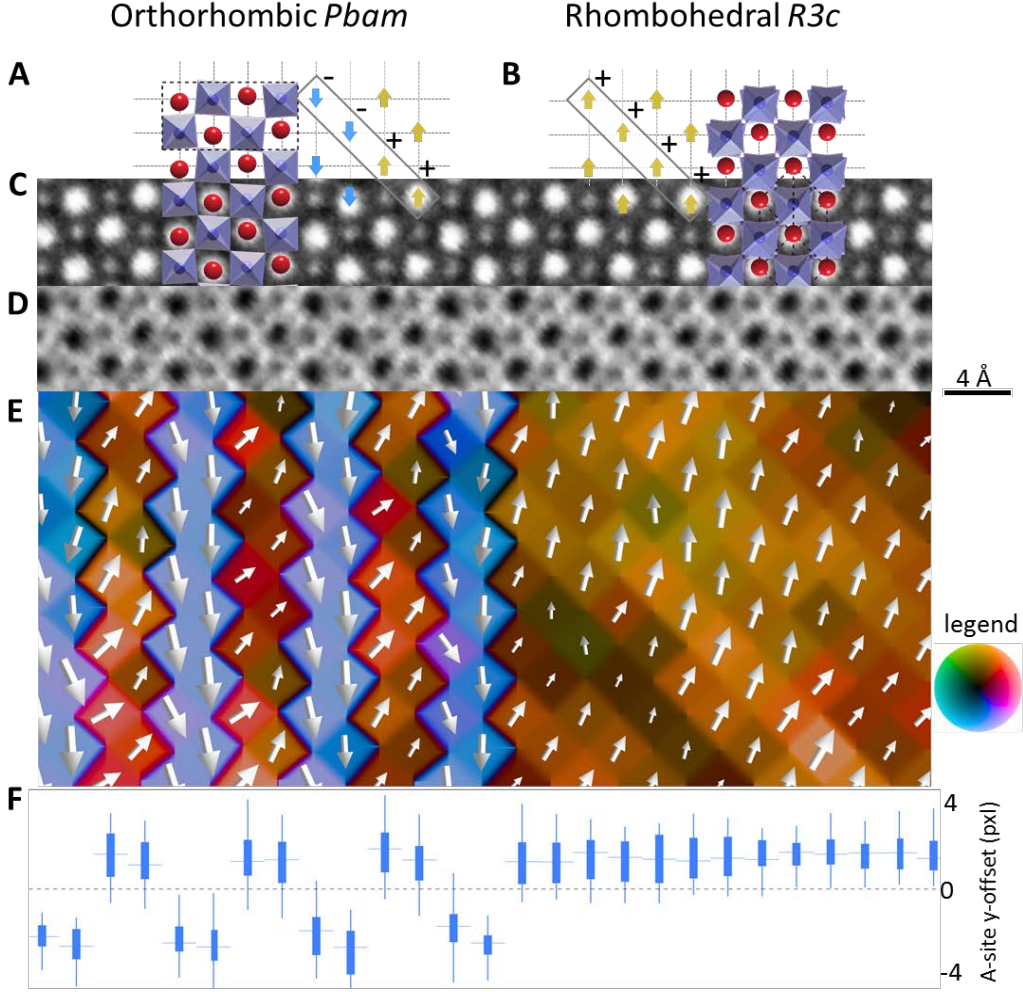

**FIGURE 1.** STEM of the $La_{0.17}Bi_{0.83}FeO_3$. The interface between the orthorhombic **O** (left) and rhombohedral **R** (right) phases is shown. **(a)** Atomic model of the O-*Pbam* antiferroelectric type structure + +− − [2]. **(b)** Atomic model of the rhombohedral ferroelectric *R3c* structure + + + + [17]. **(c)** Atomic resolution HAADF image. **(d)** Atomic resolution ABF image. **(e)** A-site displacements. **(f)**. Box and whisker plot of the [101] displacement component per layer (~35 datapoints per column).

## III. THEORETICAL DESCRIPTION

Here we describe the FSM in accordance with experimental results, shown in **Fig.1.** The conventional LGD free energy density is a sum of Landau, gradient and surface energies:

$$G = \int_V \left( G_{Landau} + G_{grad} \right) + G_S . \tag{1a}$$

Landau free energy expansion, containing the quadratic and bilinear contributions of the A-cations displacements $A_i$ ($i$=1-4) in $ABO_3$ perovskite with m3m parent phase, is:

$$G_{Landau} = \frac{\alpha}{2} A_i A_i + \mu(A_1 A_2 + A_2 A_3 + A_3 A_4 + A_1 A_4) + \eta(A_1 A_3 + A_2 A_4) + \frac{\beta}{4} A_i^2 A_i^2$$
$$+ \frac{\gamma}{2}(A_1^2 A_2^2 + A_2^2 A_3^2 + A_3^2 A_4^2 + A_1^2 A_4^2) + \frac{\delta}{2}(A_1^2 A_3^2 + A_2^2 A_4^2) + \lambda A_1 A_2 A_3 A_4 + ...$$
(1b)

Here we assume that only the first term in Eq.(1b) has temperature dependent coefficient, namely $\alpha = \alpha_T (T - T_C)$, and all constants can depend on the global or local content of impurity (e.g. La atoms).

The atomic displacements of different sub-lattices (which are equivalent in undoped ABO$_3$) could be considered as long-range order parameters, $A_1$, $A_2$, $A_3$ and $A_4$. We further assume that the standard inequality $\frac{\beta}{4} > \frac{\gamma}{2} > \frac{\delta}{2} \gg \lambda$ is valid, as necessary for the functional stability and expansion series convergence.

The gradient energy $G_{grad} = g_{ijkl} \frac{\partial A_i}{\partial x_k} \frac{\partial A_j}{\partial x_l}$ will be considered in the simplest isotropic approximation for the gradient coefficient tensor $g_{ijkl}$, at that all physical quantities are x-dependent:

$$G_{grad} = g_{ijkl} \frac{\partial A_i}{\partial x_k} \frac{\partial A_j}{\partial x_l} \approx g \left( \frac{\partial A_i}{\partial x} \right)^2.$$
(1c)

The surface energy is assumed to be a positively defined quadratic form,

$$G_S = \frac{\alpha_S}{2}(A_1^2 + A_2^2 + A_3^2 + A_4^2).$$
(1d)

Using Dzyaloshinsky substitution [28]:

$$B_1 = \frac{A_1 + A_2 + A_3 + A_4}{2}, \quad B_2 = \frac{A_1 - A_2 - A_3 + A_4}{2}, \quad B_3 = \frac{A_1 - A_2 + A_3 - A_4}{2}, \quad B_4 = \frac{A_1 + A_2 - A_3 - A_4}{2}$$
(2)

and making elementary algebraic transformation listed in **Appendix A** (see Supplementary materials [29]), one could rewrite the Eq.(1b) as follows

$$G_{Landau} = \frac{\alpha^*}{2} B_1^2 + \frac{\mu^*}{2}(B_2^2 + B_4^2) + \frac{\eta^*}{2} B_3^2 + \frac{\beta^*}{4}(B_1^4 + B_2^4 + B_3^4 + B_4^4) +$$
$$\gamma^*(B_1^2 B_2^2 + B_2^2 B_3^2 + B_3^2 B_4^2 + B_1^2 B_4^2) + \frac{\delta^*}{4}(B_2^2 B_4^2 + B_1^2 B_3^2) + \lambda^* B_1 B_2 B_3 B_4$$
(3a)

The expansion coefficients:

$$\alpha^* = \alpha + 2\mu + \eta, \quad \mu^* = \alpha - \eta, \quad \eta^* = \alpha - 2\mu + \eta, \quad \beta^* = \frac{\beta}{4} + \frac{\gamma}{2} + \frac{\delta}{4} + \frac{\lambda}{4},$$
$$\gamma^* = \frac{3\beta}{8} + \frac{\gamma}{4} - \frac{\delta}{8} - \frac{\lambda}{8}, \quad \delta^* = \frac{3\beta}{2} - \gamma + \frac{3\delta}{2} - \frac{\lambda}{2}, \quad \lambda^* = \frac{3\beta}{2} - \gamma - \frac{\delta}{2} + \frac{\lambda}{2}$$
(3b)

Under the condition $\lambda^* > 0$ expression (3a) contains several possible phase transitions from the paraelectric phase to the different homogeneous phases (R and O) or spatially modulated phases (SM I, II, III), which are listed in **Table I**.

To model the boundary between coexisting R, O and SM phases, one can solve numerically coupled Euler-Lagrange equations obtained by the variation of the free energy (3). The equations are supplemented by the third kind boundary conditions (**BCs**) steaming from the surface energy variation with respect to the cation displacements (see **Appendix A** [29] for details). Being interested in the coexistence of different phases in a thin $ABO_3$ film, we compared the limiting cases of zero BCs, $B_i\big|_{x=0,h} = 0$, with natural BCs, $\frac{\partial B_i}{\partial x}\bigg|_{x=0,h} = 0$, and conditions of the components periodicity in a bulk sample.

**Table I.** Description of different homogeneous phases in Eqs.(1)-(3) and necessary conditions of their stability

| Phase name | Signs of $A_1A_2A_3A_4$ | Values of the order parameters $B_i$ and $A_i$ | Necessary conditions and corresponding energy value |
|---|---|---|---|
| Para-phase | "0000" $----$ | $B_1 = 0$, $B_2 = 0$, $B_3 = 0$, $B_4 = 0$ <br> $A_1 = 0$, $A_2 = 0$, $A_3 = 0$, $A_4 = 0$ | $\frac{\alpha}{2} - \|\mu\| + \frac{\eta}{2} \geq 0$, $\frac{\alpha}{2} - \frac{\eta}{2} \geq 0$ <br> $G_P = 0$ |
| Homogeneous (R-phase) | "++++" | $B_1 = \sqrt{-\frac{\alpha^*}{\beta^*}}$, $B_2 = 0$, $B_3 = 0$, $B_4 = 0$ <br> $A_1 = A_2 = A_3 = A_4 = \sqrt{-\frac{\alpha^*}{4\beta^*}}$ | $\alpha + 2\mu + \eta < 0$, <br> $\beta + 2\gamma + \delta + \lambda > 0$ <br> $G_R = \frac{-\alpha^{*2}}{\beta^*}$ |
| Modulated I (O-phase) | "+--+" or "-++-" | $B_2 = \sqrt{-\frac{\mu^*}{\beta^*}}$, (or the same $B_4$) <br> $B_1 = 0$, $B_3 = 0$, $B_4 = 0$ (or $B_2 = 0$) <br> $A_1 = A_2 = -A_3 = -A_4 = \sqrt{-\frac{\mu^*}{4\beta^*}}$ | $\alpha - \eta < 0$, <br> $\beta + 2\gamma + \delta + \lambda > 0$, <br> $G_O = \frac{-\mu^{*2}}{4\beta^*}$ |
| Modulated II (AFE phase) | "+-+-" | $B_1 = 0$, $B_2 = 0$, $B_4 = 0$, $B_3 = \sqrt{-\frac{\eta^*}{\beta^*}}$, <br> $A_1 = -A_2 = A_3 = -A_4 = \sqrt{-\frac{\eta^*}{4\beta^*}}$ | $\alpha - 2\mu + \eta < 0$ <br> $\beta + 2\gamma + \delta + \lambda > 0$ <br> $G_A = \frac{-\eta^{*2}}{4\beta^*}$ |

| Modulated III (mixture of several phases) | "+0−0" | $B_2 = B_4 = \sqrt{-\dfrac{\alpha-\eta}{\beta+\delta}}$, $B_1 = 0$, $B_3 = 0$, $A_1 = -A_3 = \dfrac{1}{2}\sqrt{-\dfrac{\alpha-\eta}{\beta+\delta}}$, $A_2 = A_4 = 0$ | $\alpha-\eta < 0$, $\beta+\delta > 0$, $G_M = \dfrac{-(\alpha-\eta)^2}{4(\beta+\delta)}$ |
|---|---|---|---|

Since the coefficients $\alpha$, $\beta$, $\gamma$, $\lambda$, $\delta$, $\mu$ and $\eta$ in the stability conditions depend on the impurity content in the solid solution, the appearance of O and R phases and their coexistence can be explained. A gradient terms, or higher terms, or both can make the modulation in O and SM- phases much more complicated.

Formally R and O phases coexistence (that is observed by STEM) can be realized in the case of their energies equality. The coexistence condition, $G_R = G_O$, gives $\alpha - \eta = \alpha + 2\mu + \eta \Leftrightarrow -\eta = \mu$ per **Table I**, and the phases stability conditions are $\alpha + 2\mu + \eta < 0$, $\alpha - \eta < 0$ and $\beta + 2\gamma + \delta + \lambda > 0$.

In the case of the weak deviations from the phase equilibrium, $-\eta = \mu$, i.e. when the condition $\eta + \mu + \varsigma = 0$ takes place along with the inequality $\varsigma << \mu$, one could write the free energy (3) in the following dimensionless form.

$$G_{12} \approx -(\alpha+\mu)B_S^2 \int_V \left[ -(1-c)\frac{b_1^2}{2} - (1+c)\frac{b_2^2}{2} + \frac{b_1^4 + b_2^4}{4} + \frac{\chi}{2}b_1^2 b_2^2 + \frac{h}{2}\left(\left(\frac{\partial b_1}{\partial x}\right)^2 + \left(\frac{\partial b_2}{\partial x}\right)^2\right) \right] \quad (4)$$

where the order parameters $B_i = B_S b_i$ ($i=1,2$), the spontaneous value $B_S = 2\sqrt{-\dfrac{\alpha+2\mu+\eta}{\beta+2\gamma+\delta+\lambda}}$, dimensionless coupling constant $\chi = \left(\dfrac{3\beta}{4} + \dfrac{\gamma}{2} - \dfrac{\delta}{4} - \dfrac{\lambda}{4}\right)\left(\dfrac{\beta}{4} + \dfrac{\gamma}{2} + \dfrac{\delta}{4} + \dfrac{\lambda}{4}\right)^{-1}$ and gradient coefficient $h = B_S^2 g$ are introduced (see **Appendix A** for the details of calculations). The parameter $c \equiv \dfrac{\varsigma}{\alpha+\mu}$ is the sublattice asymmetry constant.

Thus, FSM reduces the description of R and O phases coexistence to the thermodynamic analyses of the free energy functional with three dimensionless phenomenological parameters, - asymmetry constant $c$, sublattices coupling strength $\chi$ and order parameters gradient energy coefficient $h$. R phase corresponds to $b_1 \neq 0$ and $b_2 = 0$, while O phase corresponds to $b_2 \neq 0$ and $b_1 = 0$.

To study the boundary between coexisting R and O phases, we solved numerically coupled Euler-Lagrange equations obtained by the variation of the energy (4) supplemented by the natural BCs, $\left.\frac{\partial b_i}{\partial x}\right|_{x=0,h} = 0$, and conditions of the components periodicity in a bulk sample.

Distribution of order parameters $b_1$ and $b_2$ near the boundary between R-domain (left) and O-domain (right) are shown in **Fig.2.** It is seen from **Fig.2(a)** that the increase of the dimensionless coupling constant $\chi$ leads to the narrowing of the interfacial region between R and O phases. Actually, the higher is the term $\frac{\chi}{2}b_1^2 b_2^2$, the stronger is the coupling between the dimensionless order parameters $b_1$ and $b_2$. For a weak coupling corresponding to $\chi<1$ two separate R and O phases are unstable [and the case is not shown in the **Fig. 2(a)**]. As one can see from **Fig.2(b)** the increase of the sublattice asymmetry parameter $c$ supports R-phase with $b_2=0$. In particular, the saturation value of $b_1$ decreases and tends to disappear with further increase of $c$. At the same time, the width of the interfacial R-O region is almost unaffected by the variation of the parameter $c$. Note that $c$ value can be regarded proportional to the impurity concentration, while $\chi$ and $h$ are regarded concentration independent.

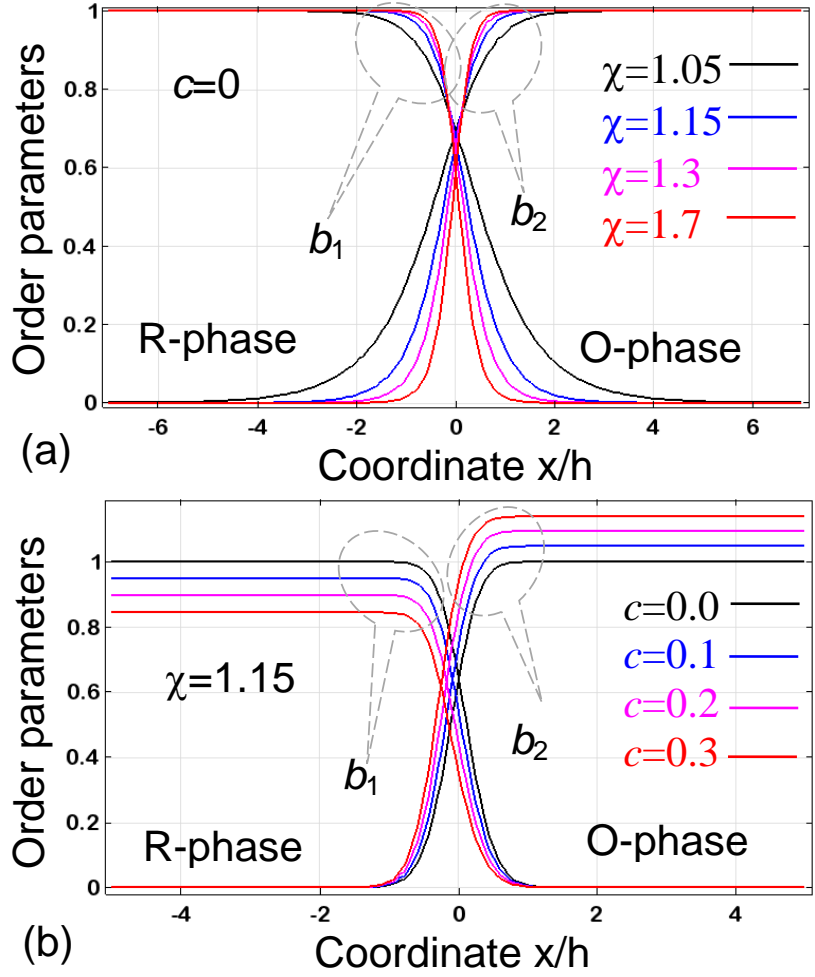

**FIGURE 2.** Distribution of order parameters $b_1$ and $b_2$ near the interface between R-phase (left) and O-phase (right) **(a)** for $c=0$ and different values of parameter $\chi=1.05$, 1.15, 1.3 and 1.7 (black, blue, magenta and red curves respectively); **(b)** for $\chi=1.15$ and different values of parameter $c=0$, 0.1, 0.2 and 0.3 (black, blue, magenta and red curves respectively).

## IV. COMPARISON WITH EXPERIMENT

Though the order parameters $B_i$ are very convenient for the theoretical description of the phase diagram, their distribution cannot be observed directly in STEM experiments, while the distribution of the order parameters $A_i$ indeed can. The distribution of the normalized order parameters $A_1$ and $A_2$ near the boundary between R-domain (left) and O-domain (right) are shown in **Fig.3.** Let us compare **Fig.3** with experimental results shown in the bottom of **Fig.1e.** As one can see the semi-quantitative agreement is present between the **Fig.3** and **Fig.1e**, because $A_1 \approx A_2 \approx A_3 \approx A_4$ in R-domain, while the value of $A_1 \approx A_2$ and $A_3 \approx A_4$ change their signs in O-domain.

Cations A displacement map near the phase boundary between R-domain and O-domain has been calculated theoretically and shown in **Fig.4.** Note the evident agreement between the theoretical **Fig.4** and experimental results shown in **Fig.1e,f**, because the contrast is absent in R-domain, where $A_1 \approx A_2 \approx A_3 \approx A_4$, while it is alternating in O-domain, where the value of $A_1 \approx A_2$ and $A_3 \approx A_4$ change their signs.

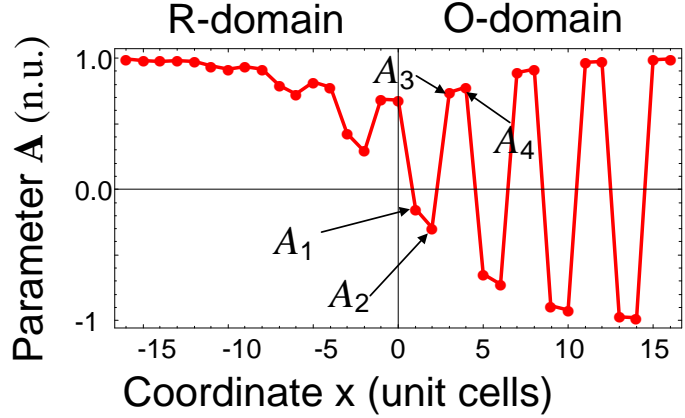

**FIGURE 3.** Distribution of the normalized order parameters $A_i$ (in cyclic order) near the boundary between R-domain (left) and O-domain (right).

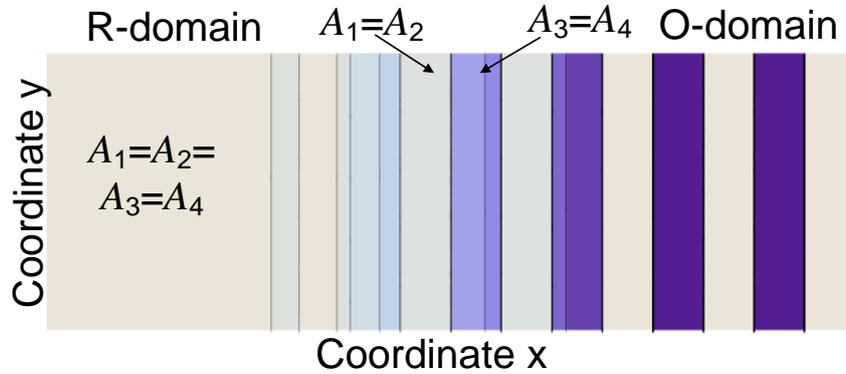

**FIGURE 4.** Cations A displacement map according to theoretical calculations near the boundary between R-domain (left) and O-domain (right).

## V. CONCLUSION

We propose a LGD-type free energy describing the displacements of A-cation sublattices in (anti)ferroelectric-antiferrodistortive perovskites of $ABO_3$-type. The four sublattices model, shortly FSM, proposes analytical description of the A-cation displacements in four neighboring cells and determines the conditions of different (O, R and SM) phases appearance and stability in pristine and doped $ABO_3$-type perovskites. Thus FSM explains the atomic displacements in La-doped BFO we observed by atomic-resolution STEM measurements.

FSM reduces the description of R and O phases coexistence to the thermodynamic analyses of the free energy functional with three dimensionless parameters, such as sublattice asymmetry constant $c$, their coupling strength $\chi$ and gradient energy coefficient $h$. Increase of the constant $\chi$ leads to the narrowing of the interface region between R and O phases. For a weak coupling between sublattices (corresponding to $\chi<1$) two separate R and O phases becomes unstable. The increase of the asymmetry parameter $c$ supports R-phase. At the same time, the width of the interfacial R-O region is almost unaffected by the variation of the parameter $c$. Note that $c$ value can be proportional to La concentration in BFO, while $\chi$ and $h$ are regarded concentration independent.

The FSM model has the advantage of deriving from a directly observable order parameter in atomic-scale STEM measurements. For the large A-site displacive type Pb- or Bi-based (anti)ferrodistortive-(anti)ferroelectrics [like $(Bi,La)FeO_3$ or $Pb(Ti,Zr)O_3$] this method also maximizes the experimental precision, as it derives exclusively from the strongest scattering atomic columns undergoing the largest displacements from their centrosymmetric positions within unit cell.

**Acknowledgements**

Authors are very grateful to the Referee for excellent suggestion how to improve the form of Dzyaloshinsky substitution. This material is based upon work (S.V.K, C.T.N.) supported by the U.S. Department of Energy, Office of Science, Office of Basic Energy Sciences, and performed in the Center for Nanophase Materials Sciences, supported by the Division of Scientific User Facilities. D.C. thanks the financial support from the National Natural Science Foundation of China (Grant Nos. U1832104 and 11704130). A.N.M work has received funding from the European Union's Horizon 2020 research and innovation programme under the Marie Skłodowska-Curie grant agreement No 778070, and partially supported by the National Academy of Sciences of Ukraine (project No. 0117U002612).

**Authors' contribution.** A.N.M. and E.A.E. proposed the theoretical model and performed calculations. D.C. performed the LBFO film PLD synthesis. C.T.N. performed and analyzed the STEM experiments. S.V.K. generated the research idea, interpreted theoretical and experimental results and wrote the manuscript draft. All co-authors worked on the results discussion and manuscript improvement.

# SUPPLEMENTARY MATERIALS

## APPENDIX A. Free energy of the system, consisting of the four sub-lattices

Quadratic and bilinear terms contributions to free energy expansion on the cation A displacement $A_i$ could be written as:

$$G = \int_V (G_{Landau} + G_{grad}) + \int_S G_S, \quad (A.1a)$$

$$G_{Landau} = \frac{\alpha}{2}(A_1^2 + A_2^2 + A_3^2 + A_4^2) + \mu(A_1 A_2 + A_2 A_3 + A_3 A_4 + A_1 A_4) + \eta(A_1 A_3 + A_2 A_4)$$
$$+ \frac{\beta}{4}(A_1^4 + A_2^4 + A_3^4 + A_4^4) + \frac{\gamma}{2}(A_1^2 A_2^2 + A_2^2 A_3^2 + A_3^2 A_4^2 + A_1^2 A_4^2) + \frac{\delta}{2}(A_1^2 A_3^2 + A_2^2 A_4^2) \quad (A.1b)$$
$$+ \lambda A_1 A_2 A_3 A_4 + \xi(A_1 A_2^3 + A_2 A_3^3 + A_3 A_4^3 + A_1 A_4^3 + ...)$$

Note that $\frac{\beta}{4} > \frac{\gamma}{2} > \frac{\delta}{2} \gg \lambda \gg \xi$, and so we can neglect $\xi$-terms.

$$G_{grad} = g_{ijkl} \frac{\partial A_i}{\partial x_k} \frac{\partial A_j}{\partial x_l} + \{\text{higher order terms}\} \rightarrow g\left(\left(\frac{\partial A_1}{\partial x}\right)^2 + \left(\frac{\partial A_2}{\partial x}\right)^2 + \left(\frac{\partial A_3}{\partial x}\right)^2 + \left(\frac{\partial A_4}{\partial x}\right)^2\right) + ...$$
(A.1c)

Here we supposed that only the first term has temperature dependent coefficient, namely $\alpha = \alpha_T(T - T_0)$, but all constants can depend on the global or local content of La. Long-range order parameters $A_1$, $A_2$, $A_3$ and $A_4$ could be considered as the atomic displacements of different sub-lattices (which are equivalent in pristine BFO). The "simplest" surface energy is assumed to be a positively defined quadratic form, $G_S = \frac{\alpha_S}{2}(A_1^2 + A_2^2 + A_3^2 + A_4^2)$.

Using Dzyaloshinsky substitution

$$B_1 = \frac{A_1 + A_2 + A_3 + A_4}{2}, \quad B_2 = \frac{A_1 - A_2 - A_3 + A_4}{2}, \quad (A.2a)$$

$$B_3 = \frac{A_1 - A_2 + A_3 - A_4}{2}, \quad B_4 = \frac{A_1 + A_2 - A_3 - A_4}{2}, \quad (A.2b)$$

One could rewrite the homogeneous part of Eq.(A.1b) as follows

$$G_{Landau} = \frac{\alpha}{2}\left(B_1^2 + B_2^2 + B_3^2 + B_4^2\right) + \mu\left(B_1^2 - B_3^2\right) + \frac{\eta}{2}\left(B_1^2 - B_2^2 + B_3^2 - B_4^2\right)$$
$$+ \frac{\beta}{4}\left(\frac{B_1^4 + B_2^4 + B_3^4 + B_4^4}{4} + \frac{3}{2}\left(B_1^2 B_2^2 + B_2^2 B_3^2 + B_3^2 B_4^2 + B_1^2 B_4^2 + B_2^2 B_4^2 + B_1^2 B_3^2\right) + 6 B_1 B_2 B_3 B_4\right) +$$
$$\frac{\gamma}{2}\left(\frac{B_1^4 + B_2^4 + B_3^4 + B_4^4}{4} + \frac{1}{2}\left(B_1^2 B_2^2 + B_2^2 B_3^2 + B_3^2 B_4^2 + B_1^2 B_4^2 - B_2^2 B_4^2 - B_1^2 B_3^2\right) - 2 B_1 B_2 B_3 B_4\right) +$$
$$\frac{\delta}{2}\left(\frac{B_1^4 + B_2^4 + B_3^4 + B_4^4}{8} - \frac{1}{4}\left(B_1^2 B_2^2 + B_2^2 B_3^2 + B_3^2 B_4^2 + B_1^2 B_4^2 - 3 B_2^2 B_4^2 - 3 B_1^2 B_3^2\right) - B_1 B_2 B_3 B_4\right) +$$
$$\lambda\left(\frac{B_1^4 + B_2^4 + B_3^4 + B_4^4}{16} - \frac{1}{8}\left(B_1^2 B_2^2 + B_2^2 B_3^2 + B_3^2 B_4^2 + B_1^2 B_4^2 + B_2^2 B_4^2 + B_1^2 B_3^2\right) + \frac{1}{2} B_1 B_2 B_3 B_4\right)$$

(A.3a)

After elementary algebraic transformations Eq.(A.3a) can be rewritten as

$$G_{Landau} = \begin{bmatrix} (\alpha + 2\mu + \eta)\frac{B_1^2}{2} + (\alpha - \eta)\frac{B_2^2 + B_4^2}{2} + (\alpha - 2\mu + \eta)\frac{B_3^2}{2} \\ + \left(\frac{\beta}{4} + \frac{\gamma}{2} + \frac{\delta}{4} + \frac{\lambda}{4}\right)\frac{B_1^4 + B_2^4 + B_3^4 + B_4^4}{4} \\ + \left(\frac{3\beta}{8} + \frac{\gamma}{4} - \frac{\delta}{8} - \frac{\lambda}{8}\right)\left(B_1^2 B_2^2 + B_2^2 B_3^2 + B_3^2 B_4^2 + B_1^2 B_4^2\right) \\ + \left(\frac{3\beta}{2} - \gamma + \frac{3\delta}{2} - \frac{\lambda}{2}\right)\frac{B_2^2 B_4^2 + B_1^2 B_3^2}{4} + \left(\frac{3\beta}{2} - \gamma - \frac{\delta}{2} + \frac{\lambda}{2}\right) B_1 B_2 B_3 B_4 \end{bmatrix}$$

(A.3b)

Under the condition $\frac{3\beta}{2} - \gamma - \frac{\delta}{2} + \frac{\lambda}{2} > 0$ expression (A.3b) contains several possible phase transitions from para-phase to the different homogeneous and modulated phases, which are summarized in **Table I**.

**Table I.** Description of different homogeneous phases of the system and the necessary conditions of their stability

| Phase name | Signs and arrows of $A_1 A_2 A_3 A_4$ | Values of $B_i$ and $A_i$ | Necessary conditions and corresponding energy value |
|---|---|---|---|
| Para-phase | "0000" <br> − − − − | $B_1 = 0$, $B_2 = 0$, $B_3 = 0$, $B_4 = 0$ <br> $A_1 = 0$, $A_2 = 0$, $A_3 = 0$, $A_4 = 0$ | $\frac{\alpha}{2} - |\mu| + \frac{\eta}{2} \geq 0$, <br> $\frac{\alpha}{2} - \frac{\eta}{2} \geq 0$ <br> $G_P = 0$ |

| | | | |
|---|---|---|---|
| Homogeneous (R-domain) | "++++" ↑↑↑↑ | $B_1 = 2\sqrt{-\dfrac{\alpha + 2\mu + \eta}{\beta + 2\gamma + \delta + \lambda}}$, $B_2 = 0$, $B_3 = 0$, $B_4 = 0$ $A_1 = A_2 = A_3 = A_4 = \sqrt{-\dfrac{\alpha + 2\mu + \eta}{\beta + 2\gamma + \delta + \lambda}}$ | $\alpha + 2\mu + \eta < 0$, $\beta + 2\gamma + \delta + \lambda > 0$ $G_R = \dfrac{-(\alpha + 2\mu + \eta)^2}{\beta + 2\gamma + \delta + \lambda}$ |
| Modulated I (O-domain) | "+--+" Or "-++-" ↑↓↓↑ ↓↑↑↓ | $B_2 = 2\sqrt{-\dfrac{\alpha - \eta}{\beta + 2\gamma + \delta + \lambda}}$, (or the same $B_4$) $B_1 = 0$, $B_3 = 0$, $B_4 = 0$ (or $B_2 = 0$) $A_1 = A_2 = -A_3 = -A_4 = \sqrt{-\dfrac{\alpha - \eta}{\beta + 2\gamma + \delta + \lambda}}$ | $\alpha - \eta < 0$, $\beta + 2\gamma + \delta + \lambda > 0$, $G_O = \dfrac{-(\alpha - \eta)^2}{\beta + 2\gamma + \delta + \lambda}$ |
| Modulated II (AFE phase) | "+-+-" ↑↓↑↓ | $B_1 = 0$, $B_2 = 0$, $B_4 = 0$ $B_3 = 2\sqrt{-\dfrac{\alpha - 2\mu + \eta}{\beta + 2\gamma + \delta + \lambda}}$, $A_1 = -A_2 = A_3 = -A_4 = \sqrt{-\dfrac{\alpha - 2\mu + \eta}{\beta + 2\gamma + \delta + \lambda}}$ | $\alpha - 2\mu + \eta < 0$ $\beta + 2\gamma + \delta + \lambda > 0$ $G_A = \dfrac{-(\alpha - 2\mu + \eta)^2}{\beta + 2\gamma + \delta + \lambda}$ |
| Modulated III (possible mixture of several phases) | "+0-0" ↑-↓-↑-↓- | $B_2 = B_4 = \sqrt{-\dfrac{\alpha - \eta}{\beta + \delta}}$, $B_1 = 0$, $B_3 = 0$, $A_1 = -A_3 = \dfrac{1}{2}\sqrt{-\dfrac{\alpha - \eta}{\beta + \delta}}$, $A_2 = A_4 = 0$ | $\alpha - \eta < 0$, $\beta + \delta > 0$, $G_M = \dfrac{-(\alpha - \eta)^2}{4(\beta + \delta)}$ |

Since the coefficients α, β, γ, λ, δ, μ and η in the stability conditions depends on the La-content in the solid solution, the appearance of O- and R-domains and their coexistence can be explained. A gradient terms, or higher terms, or both can make the modulation in o- and m- phases much more complicated.

Formally r-o-m domains coexistence can be realized the case of their energies equality. Corresponding Euler-Lagrange equations are:

$$(\alpha + 2\mu + \eta)B_1 + \left(\frac{\beta}{4} + \frac{\gamma}{2} + \frac{\delta}{4} + \frac{\lambda}{4}\right)B_1^3 + \left(\frac{3\beta}{4} + \frac{\gamma}{2} - \frac{\delta}{4} - \frac{\lambda}{4}\right)B_1\left(B_2^2 + B_4^2\right) +$$
$$\left(\frac{3\beta}{4} - \frac{\gamma}{2} + \frac{3\delta}{4} - \frac{\lambda}{4}\right)B_1 B_3^2 + \left(\frac{3\beta}{2} - \gamma - \frac{\delta}{2} + \frac{\lambda}{2}\right)B_2 B_3 B_4 - g\frac{d^2}{dx^2}B_1 = 0 \qquad (A.4a)$$

$$(\alpha - \eta)B_2 + \left(\frac{\beta}{4} + \frac{\gamma}{2} + \frac{\delta}{4} + \frac{\lambda}{4}\right)B_2^3 + \left(\frac{3\beta}{4} + \frac{\gamma}{2} - \frac{\delta}{4} - \frac{\lambda}{4}\right)B_2\left(B_1^2 + B_3^2\right) +$$
$$\left(\frac{3\beta}{4} - \frac{\gamma}{2} + \frac{3\delta}{4} - \frac{\lambda}{4}\right)B_2 B_4^2 + \left(\frac{3\beta}{2} - \gamma - \frac{\delta}{2} + \frac{\lambda}{2}\right)B_1 B_3 B_4 - g\frac{d^2}{dx^2}B_2 = 0 \quad \text{(A.4b)}$$

$$(\alpha - 2\mu + \eta)B_3 + \left(\frac{\beta}{4} + \frac{\gamma}{2} + \frac{\delta}{4} + \frac{\lambda}{4}\right)B_3^3 + \left(\frac{3\beta}{4} + \frac{\gamma}{2} - \frac{\delta}{4} - \frac{\lambda}{4}\right)B_3\left(B_2^2 + B_4^2\right) +$$
$$\left(\frac{3\beta}{4} - \frac{\gamma}{2} + \frac{3\delta}{4} - \frac{\lambda}{4}\right)B_3 B_1^2 + \left(\frac{3\beta}{2} - \gamma - \frac{\delta}{2} + \frac{\lambda}{2}\right)B_1 B_2 B_4 - g\frac{d^2}{dx^2}B_3 = 0 \quad \text{(A.4c)}$$

$$(\alpha - \eta)B_4 + \left(\frac{\beta}{4} + \frac{\gamma}{2} + \frac{\delta}{4} + \frac{\lambda}{4}\right)B_4^3 + \left(\frac{3\beta}{4} + \frac{\gamma}{2} - \frac{\delta}{4} - \frac{\lambda}{4}\right)B_4\left(B_1^2 + B_3^2\right) +$$
$$\left(\frac{3\beta}{4} - \frac{\gamma}{2} + \frac{3\delta}{4} - \frac{\lambda}{4}\right)B_4 B_2^2 + \left(\frac{3\beta}{2} - \gamma - \frac{\delta}{2} + \frac{\lambda}{2}\right)B_1 B_2 B_3 - g\frac{d^2}{dx^2}B_4 = 0 \quad \text{(A.4d)}$$

Equations (4) should be supplemented by the third kind boundary conditions (**BCs**) steaming from the surface energy variation with respect to the cation displacements. Being interested in the coexistence of different phases in a thin BFO film, let us compare the limiting cases of zero conditions, $B_i\big|_{x=0,h} = 0$, with natural conditions, $\frac{\partial B_i}{\partial x}\bigg|_{x=0,h} = 0$, and conditions of the components periodicity in a bulk sample.

Specifically, the coupled equations for R- and O-phases coexistence (that is observed by STEM) have the form

$$(\alpha + 2\mu + \eta)B_1 + \left(\frac{\beta}{4} + \frac{\gamma}{2} + \frac{\delta}{4} + \frac{\lambda}{4}\right)B_1^3 + \left(\frac{3\beta}{4} + \frac{\gamma}{2} - \frac{\delta}{4} - \frac{\lambda}{4}\right)B_1 B_2^2 - g\frac{d^2}{dx^2}B_1 = 0, \quad \text{(A.5a)}$$

$$(\alpha - \eta)B_2 + \left(\frac{\beta}{4} + \frac{\gamma}{2} + \frac{\delta}{4} + \frac{\lambda}{4}\right)B_2^3 + \left(\frac{3\beta}{4} + \frac{\gamma}{2} - \frac{\delta}{4} - \frac{\lambda}{4}\right)B_1^2 B_2 - g\frac{d^2}{dx^2}B_2 = 0, \quad \text{(A.5b)}$$

along with the BCs $B_i\big|_{x=0,h} = 0$ and the first integral

$$I_{Landau} = \begin{bmatrix} (\alpha + 2\mu + \eta)\frac{B_1^2}{2} + (\alpha - \eta)\frac{B_2^2}{2} + \left(\frac{\beta}{4} + \frac{\gamma}{2} + \frac{\delta}{4} + \frac{\lambda}{4}\right)\frac{B_1^4 + B_2^4}{4} \\ + \left(\frac{3\beta}{8} + \frac{\gamma}{4} - \frac{\delta}{8} - \frac{\lambda}{8}\right)B_1^2 B_2^2 - \frac{g}{2}\left(\left(\frac{\partial B_1}{\partial x}\right)^2 + \left(\frac{\partial B_2}{\partial x}\right)^2\right) \end{bmatrix} \quad \text{(A.6a)}$$

And reduced free energy functional

$$G_{12} = \int_V \begin{bmatrix} (\alpha + 2\mu + \eta)\frac{B_1^2}{2} + (\alpha - \eta)\frac{B_2^2}{2} + \left(\frac{\beta}{4} + \frac{\gamma}{2} + \frac{\delta}{4} + \frac{\lambda}{4}\right)\frac{B_1^4 + B_2^4}{4} \\ + \left(\frac{3\beta}{8} + \frac{\gamma}{4} - \frac{\delta}{8} - \frac{\lambda}{8}\right)B_1^2 B_2^2 + \frac{g}{2}\left(\left(\frac{\partial B_1}{\partial x}\right)^2 + \left(\frac{\partial B_2}{\partial x}\right)^2\right) \end{bmatrix}, \quad \text{(A.6b)}$$

For which the coexistence condition, $G_R = G_O$, that gives $\alpha - \eta = \alpha + 2\mu + \eta \Leftrightarrow -\eta = \mu$ per **Table I**, and the phases stability conditions $\alpha + 2\mu + \eta < 0$, $\alpha - \eta < 0$ and $\beta + 2\gamma + \delta + \lambda > 0$. Under the conditions $-\eta = \mu$ free energy (A.6b) can be rewritten in dimensionless variables as

$$G_{12} = -(\alpha - \eta)B_S^2 \int_V \left[ -\frac{b_1^2}{2} - \frac{b_2^2}{2} + \frac{b_1^4 + b_2^4}{4} + \frac{\chi}{2}b_1^2 b_2^2 + \frac{h}{2}\left( \left(\frac{\partial b_1}{\partial x}\right)^2 + \left(\frac{\partial b_2}{\partial x}\right)^2 \right) \right], \quad (A.7)$$

where the $B_i = B_S b_i$ ($i$=1,2), spontaneous order parameter $B_S = 2\sqrt{-\frac{\alpha + 2\mu + \eta}{\beta + 2\gamma + \delta + \lambda}}$, dimensionless coupling constant $\chi = \left(\frac{3\beta}{4} + \frac{\gamma}{2} - \frac{\delta}{4} - \frac{\lambda}{4}\right)\left(\frac{\beta}{4} + \frac{\gamma}{2} + \frac{\delta}{4} + \frac{\lambda}{4}\right)^{-1}$ and gradient coefficient $h = B_S^2 g$ are introduced.